\documentclass[graybox]{svmult}
\usepackage{type1cm}        
\usepackage{makeidx}         
\usepackage{graphicx}        
\usepackage{multicol}        
\usepackage[bottom]{footmisc}

\usepackage{newtxtext}       %
\usepackage[varvw]{newtxmath}       

\makeindex            

\usepackage{hyperref}  
\def\R{\mathbb{R}}
\renewcommand{\vec}[1]{\mathbf{#1}}
\usepackage{subfig}
\usepackage{tikz}
\usepackage{pgfplots}
\pgfplotsset{compat=1.18}
\usepgfplotslibrary{groupplots}
\usepackage{wrapfig}

\usetikzlibrary{external}
\pgfplotsset{
  myaxis/.style={
    grid=both,
    grid style={opacity=.2},
    tick align=outside,
    tick style={black},
    xlabel near ticks,
    ylabel near ticks,
    ymode=log,
    yminorgrids=true,
    log basis y=10,
    yminorticks=true,
    legend cell align=left,
    legend columns=3,
    legend style={/tikz/every even column/.append style={column sep=6pt}, draw=none, font=\small},
    cycle list name=color list,
    every axis plot/.append style={line width=0.8pt, mark size=1.5pt},
  },
}

\begin{document}

\title*{Adaptive Multidimensional Quadrature on Multi-GPU Systems}

\author{Melanie Tonarelli\orcidID{0009-0005-1264-8159}, Simone Riva\orcidID{0000-0003-1032-0609},\\ Pietro Benedusi\orcidID{0000-0001-7799-5999}, Fabrizio Ferrandi\orcidID{0000-0003-0301-4419},\\ and Rolf Krause\orcidID{0000-0001-5408-5271}}

\authorrunning{M. Tonarelli et al.}

\institute{
  \textit{Melanie Tonarelli} \at
  Euler Institute, Faculty of Informatics, Università della Svizzera Italiana, Lugano, Switzerland, 
  and Politecnico di Milano, Milan, Italy,
  \email{melanie.tonarelli@usi.ch}
  \and
  \textit{Pietro Benedusi} \and \textit{Simone Riva} \at
  Istituto ricerche solari Aldo e Cele Daccò (IRSOL), Faculty of Informatics, 
  Università della Svizzera Italiana, Locarno, Switzerland, 
  \email{\{pietro.benedusi, simone.riva\}@irsol.usi.ch}
  \and
  \textit{Fabrizio Ferrandi} \at
  Politecnico di Milano, Milan, Italy,
  \email{fabrizio.ferrandi@polimi.it}
  \and
  \textit{Rolf Krause} \at
  AMCS, KAUST, King Abdullah University of Science and Technology, Thuwal, Saudi Arabia,  
  Euler Institute, Faculty of Informatics, Università della Svizzera Italiana, Lugano, Switzerland,  \email{rolf.krause@kaust.edu.sa}
}

\maketitle
\abstract{We introduce a distributed adaptive quadrature method that formulates multidimensional integration as a hierarchical domain decomposition problem on multi-GPU architectures. 
The integration domain is recursively partitioned into subdomains whose refinement is guided by local error estimators. 
Each subdomain evolves independently on a GPU, which exposes a significant load imbalance as the adaptive process progresses. 
To address this challenge, we introduce a decentralised load redistribution schemes based on a cyclic round-robin policy. 
This strategy dynamically rebalance subdomains across devices through non-blocking, CUDA-aware MPI communication that overlaps with computation. 
The proposed strategy has two main advantages compared to a state-of-the-art GPU-tailored package: higher efficiency in high dimensions; and improved robustness w.r.t the integrand regularity and the target accuracy.
}

\section{Introduction}
\label{sec:introduction}
The accurate and efficient evaluation of multidimensional integrals plays a fundamental role in many computational problems; for example, it is crucial in radiative transfer \cite{benedusi2023scalable,riva2023assessment,riva2025} and probabilistic design \cite{probabilistic-design} applications. 
Among the available approaches, deterministic adaptive quadrature methods are particularly attractive because they provide rigorous error control through recursive refinement of the integration domain. These methods iteratively subdivide the domain into smaller subregions according to local error estimators, focussing computational effort where the integrand exhibits sharp variations or singularities. 
However, in distributed systems, adaptivity can introduce severe load imbalance.
Subdomains where the integrand has a simple behaviour are computationally lightweight, whereas those featuring localised peaks or singularities require extensive refinement, resulting in uneven loads and limiting scalability. Efficient integration therefore demands distributed adaptivity and dynamic workload redistribution across subdomains.

Existing frameworks such as QUADPACK \cite{quadpack} and Cuba \cite{cuba} implement robust adaptive quadrature on CPU, and PAGANI \cite{pagani} extends this capability to a single GPU.
We introduce a distributed adaptive strategy that extends deterministic quadrature to multi-GPU architectures through dynamic domain decomposition and load redistribution.
As showcased in the experimental section, the presented strategy is efficient, robust, and accurate.

\section{Adaptive quadrature methods}
\label{sec:background}
Given a hyper-rectangle $\Omega\subset\R^d$ and $f:\Omega\to \R$ an integrable function, a \textit{quadrature rule} approximates an integral as a weighted sum of function values, i.e.
\[
 \int_{\Omega} f(\vec{x})\,\text{d}\vec{x} \; \approx \; \sum_{i = 1}^n  w_i f(\vec{x}_i), 
\]
where $\{(w_i, \vec{x}_i)\}_{i=1}^n$ are the weights and nodes of the rule.
The accuracy of a rule depends on its order and on the smoothness of the integrand. The central rule used in this work is the 9-order \textit{Genz–Malik} (GM) \cite{genz-malik}, a fully symmetric rule (nodes are distributed preserving hypercube symmetry through coordinate permutations and sign changes), with the number of nodes scaling as $\mathcal{O}(2^d)$, making it a practical choice for high-dimensional problems.

\textit{Adaptive quadrature methods} approximate multidimensional integrals by repeatedly applying a quadrature rule to progressively refined subregions of the integration domain. 
At each step, every subregion yields a local integral estimate and an associated error estimate.
Traditional adaptive algorithms maintain a prioritised list (e.g. a heap or stack) of subregions and, at each iteration, select the single subregion with the largest estimated error for refinement. That subregion is subdivided, and the process continues until the global error estimate falls below the desired tolerance.
This strategy, also known as $h$-adaptivity, concentrates computational effort where $f$ is highly oscillatory, peaked, or discontinuous.
The effectiveness of adaptive methods strongly depends on the error estimator. 
We use the heuristic estimator proposed in \cite{genz-berntsen}, which is tailored to the GM rule, to balance accuracy with computational efficiency.
Another important design choice in multidimensional refinement is how to select the axis along which to refine a subregion. 
As a common heuristic \cite{cuba}, we estimate the fourth derivative of $f$ along each axis; we then divide the subdomain along the direction with the largest estimate, where the greatest error reduction is expected.

\section{Methodology}
\label{sec:methods}
\begin{figure}[htb]
  \centering
  \setlength{\tabcolsep}{0pt}
  \begin{tabular}{cc}
    \subfloat[Single-GPU solver (adapted from \cite{pagani}).\label{fig:workflow-single}]{
      \begin{minipage}[b]{0.47\textwidth}\centering
        \includegraphics[width=0.80\linewidth]{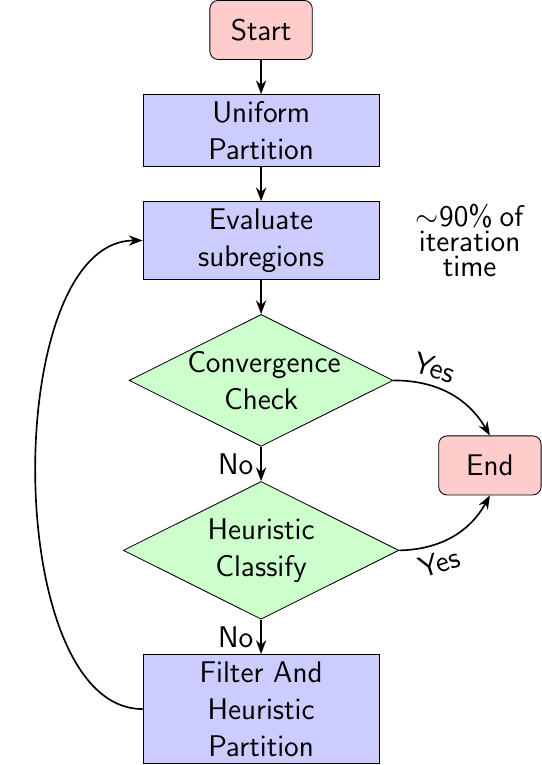}
      \end{minipage}
    }
    &
    \subfloat[Multi-GPU solver.\label{fig:workflow-multi}]{
      \begin{minipage}[b]{0.47\textwidth}\centering
        \includegraphics[width=\linewidth]{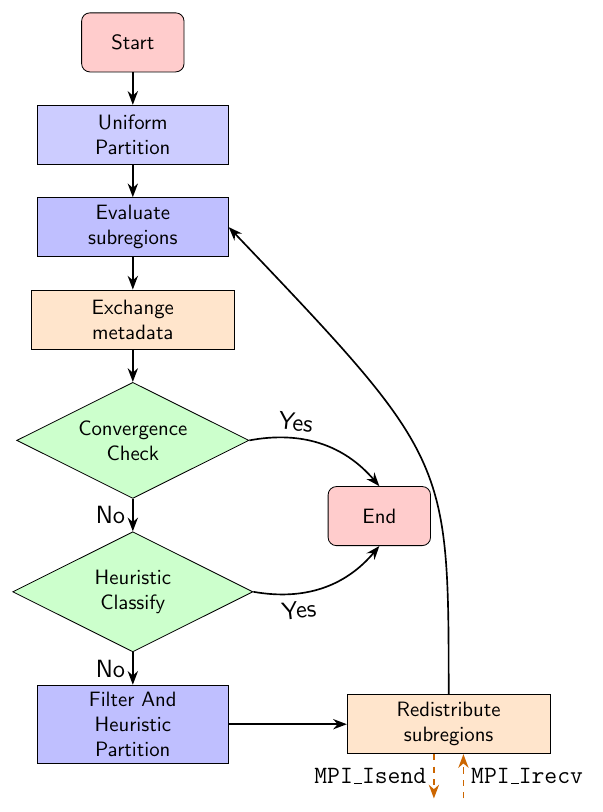}
      \end{minipage}
    }
  \end{tabular}
  \caption{Workflows of the single- and multi-GPU solvers.}
  \label{fig:workflow}
\end{figure}

The single-GPU workflow, based on that introduced in \cite{pagani}, is summarised in Figure~\ref{fig:workflow-single}. 
In contrast to traditional adaptive schemes, all subregions with a non-negligible contribution to the global error estimate are refined at each iteration. 
This strategy avoids sequential bottlenecks typical of traditional methods and is therefore better suited for GPU architectures.
The process begins with an initial uniform partition of the integration domain and proceeds iteratively: for each subdomain, the integrand is evaluated to obtain the local integral and error estimates, and the corresponding splitting axis. Then, global convergence is checked against a prescribed tolerance. A heuristic classifier discards subregions whose contribution to the error is negligible, whereas the remaining ones are subdivided along their assigned axis. The resulting subregions form the input for the next iteration, and the procedure is repeated until the convergence is reached.
All subregion data remain resident on the device throughout the computation, avoiding costly transfers between host and device. Moreover, data are stored in a Structure-of-Arrays (SoA) layout, ensuring coalesced memory accesses and efficient use of GPU memory bandwidth.
Building on this baseline workflow, we introduced modifications aimed at improving robustness and generality. This framework accommodates multiple quadrature rules, including a tensor-product Gauss–Kronrod rule \cite{Kronrod1965} currently limited to a single GPU. We use numerical guards following \cite{gander-gautschi} to mitigate round-off errors and singularities, ensuring stable convergence and preventing over-refinement. Furthermore, the filtering and splitting stages are fused into a single GPU kernel to reduce data movement and increase computational efficiency.

Extending adaptive quadrature from one GPU to many is not straightforward.
A naive strategy would be to partition the integration domain uniformly across devices and let each GPU run the adaptive algorithm independently. However, depending on the behaviour of the integrand in the subdomain, some GPUs may terminate quickly while others remain overloaded. This imbalance is the main challenge of the multi-GPU extension and motivates the need for dynamic workload redistribution.
We extended the single-GPU workflow to a distributed-memory setting using the Message Passing Interface (MPI).
At first, the global integration domain is uniformly partitioned into more subdomains than there are devices, and these are distributed across ranks so that each GPU begins with several regions to process. 
This reduces the risk that one GPU is assigned only a particularly difficult region while others finish early.
Once the initial distribution is complete, each GPU executes the workflow depicted in Figure~\ref{fig:workflow-multi}, which extends the single-GPU workflow with two additional steps:
(i) \textit{redistribution}: at the end of each iteration, the subregions are redistributed to improve load balance. Communication occurs after splitting (transferring subregion coordinates rather than full data structures) using CUDA-aware, non-blocking MPI primitives. Transfers are overlapped with GPU computation to hide latency; 
(ii) \textit{exchange of metadata}: after each `evaluation' step, devices exchange compact metadata summarising their local progress through MPI collectives. Each record includes the partial integral and error contributions, as well as in-flight estimates that conservatively bound the contribution of subregions currently in transit, 
ensuring correct convergence detection.
This step represents the only global synchronisation point of the algorithm.
A \textit{redistribution policy} specifies (i) how \textit{donors}, i.e.,\ GPUs holding more subregions than the global fair share, are paired with \textit{receivers}, i.e.,\ GPUs holding fewer, and (ii) how many subregions are transferred.
Transfers are constrained by a communication cap that limits the number of subregions per message. Within this bound, donors select a small batch of subregions with the largest error estimates, chosen after sorting, since these regions are most likely to require further refinement and therefore provide useful work for the receivers.
Currently, the strategy relies on a \textit{round-robin} policy, in which the devices are paired according to a cyclical schedule.
The scheme is lightweight, deterministic, conflict-free and guarantees that, over successive rounds, each GPU is paired with every other GPU. 
Its main limitation is that when two donors or two receivers are paired, no transfer occurs in that round, which reduces the effectiveness of load balancing. 
Moreover, the partner sequence is not topology-aware, so some exchanges may cross nodes unnecessarily, even when the imbalance could be resolved locally.

\section{Results}
\label{sec:results}
All experiments were carried out on the Daint Alps  supercomputer at CSCS, using NVIDIA GH200 superchips with CUDA-aware MPI. Integrands vary in dimensions, smoothness, and difficulty, ensuring a complete stress test:
\begingroup
\allowdisplaybreaks
\[
\setlength{\arraycolsep}{3pt}
\begin{array}{r@{\,}l r@{\,}l}
f_1(\vec{x}) &= \cos\!\left(\sum_{i=1}^d i x_i\right), &
f_2(\vec{x}) &= \displaystyle\prod_{i=1}^{d} \frac{1}{50^{-2} + (x_i-\tfrac12)^2},\\[2pt]
f_3(\vec{x}) &= (1+\sum_{i=1}^d i x_i)^{-(d+1)}, &
f_4(\vec{x}) &= \exp\!\left(-25^2\sum_{i=1}^d(x_i-\tfrac12)^2\right),\\[2pt]
f_5(\vec{x}) &= \exp\!\left(-10\sum_{i=1}^d|x_i-\tfrac12|\right), &
f_7(\vec{x}) &= \left(\sum_{i=1}^d x_i^2\right)^{11},\\[6pt]
\multicolumn{4}{l}{
f_6(\vec{x}) =
\begin{cases}
0, & x_i > (3+i)/10,\;  i = 1,\ldots,d,\\
\exp\!\left(\sum_{i=1}^d (i+4)x_i\right), & \text{otherwise.}
\end{cases}}
\end{array}
\]
\endgroup
Unless otherwise specified, the integration domain is the hypercube $[0,1]^d$, with stopping criterion $\varepsilon \leq \max({10^{-16}, |I| \cdot \tau_{\text{rel}} )}$ with $\tau_{\text{rel}}$ a relative tolerance and $\varepsilon$ and $I$ the global error and integral estimates, respectively. 

On a single GPU, we evaluated our Genz–Malik (GM) solver and compared it with the state-of-the-art PAGANI framework \cite{pagani}.
\begin{figure}[htbp!]
  \centering
  \subfloat[Execution times of GM and PAGANI.\label{fig:pagani-time}]{
    \includegraphics[width=0.9\linewidth]{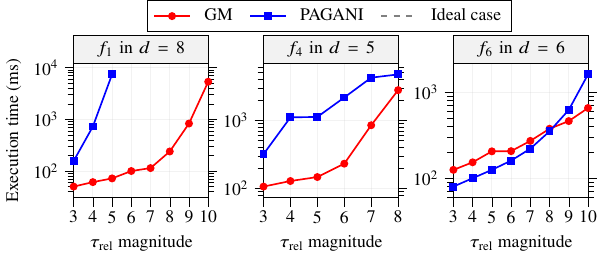}
  }\hfill
  \subfloat[Relative errors $\varepsilon_{\text{rel}}=|I-I_{\text{exact}}|/|I_{\text{exact}}|$ of GM and PAGANI, with $I_{\text{exact}}$ the exact value.\label{fig:pagani-acc}]{
    \includegraphics[width=0.95\linewidth]{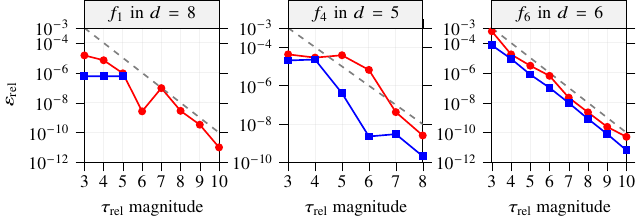}
  }
  \caption{Comparison between GM and PAGANI as a function of the prescribed tolerance $\tau_{\text{rel}} = 10^{-k}$ ($x$-axis showing $k$) on a single GPU.}
  \label{fig:pagani}
\end{figure}
Figure~\ref{fig:pagani-time} reports the execution times as a function of $\tau_{\text{rel}}$ for representative cases.
Across these benchmarks, our strategy maintained competitive or superior runtimes. Moreover, for $f_1$ (highly oscillatory), it converged for all cases, while PAGANI failed for high accuracy.
For $f_4$ (Gaussian), our strategy was faster in all tolerance ranges. For $f_6$ (discontinuous), runtimes were similar for low accuracy, while our strategy became faster when increasing accuracy.
In the remaining test functions (see Chapter 4 of \cite{Tonarelli2025Adaptive}), the picture was mixed: PAGANI was generally more efficient or comparable in peaked integrands ($f_2$, $f_3$), where its aggressive pruning avoided redundant work, while our strategy proved to be more robust in discontinuous and oscillatory functions, becoming increasingly advantageous as the tolerance tightened.
The accuracy results in Figure~\ref{fig:pagani-acc} confirm that both solvers generally met the prescribed tolerances. For $f_1$, our strategy continued to refine below $10^{-6}$ where PAGANI stalled, while for $f_4$ it occasionally overshot the target at intermediate tolerances due to overoptimistic pruning in the Gaussian tails. For all other cases, the requested accuracy was reliably achieved.
For completeness, we note that the tensor-product Gauss-Kronrod variant exhibited strong accuracy and competitive performance in low to moderate dimensions, but its computational cost increased prohibitively with $d\geq 7$.

\begin{figure}[htbp]
  \centering
  \subfloat[Feasibility comparison.\label{fig:feasibility}]{
    \includegraphics[width=0.55\linewidth]{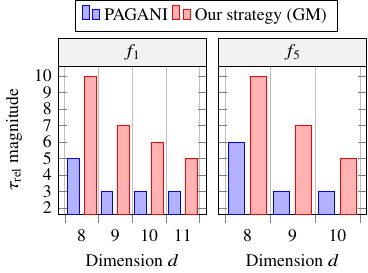}
  }\hfill
  \subfloat[Speedup.\label{fig:speedup}]{
    \includegraphics[width=0.4\linewidth]{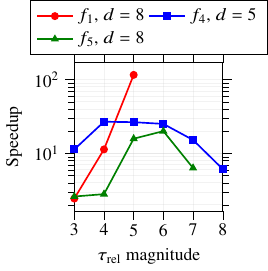}
  }
  \caption{
  Comparison between PAGANI on a single GPU and our strategy (GM) on two GPUs.
  (a) Feasibility comparison for test functions $f_1$ and $f_5$ across different dimensions. The bars indicate the strictest relative tolerances at which convergence was achieved. (b) Speedup of GM w.r.t. PAGANI.}
  \label{fig:feasibility-speedup}
\end{figure}

In the multi-GPU configuration, each MPI rank manages a device. 
At startup, the integration domain was uniformly partitioned into 8 subdomains per rank. 
By default, we used a message limit of 512 subregions.
The primary advantage of the multi-GPU extension is its ability to overcome the memory limitations of a single device. By distributing the workload, the solver is able to integrate functions at higher accuracies and dimensions, up to $d=11$ (see Figure \ref{fig:feasibility}). 
This shows that multi-GPU execution is not only a matter of performance improvement but a prerequisite for addressing certain classes of problems.
\begin{figure}[htbp]
  \centering
  \subfloat[Strong scaling.\label{fig:scaling}]{
 \includegraphics[width=0.55\linewidth]{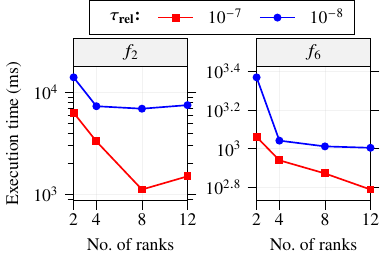}
  }\hfill
  \subfloat[Computation vs.\ idle time.\label{fig:idle-time}]{
 \includegraphics[width=0.36\linewidth]{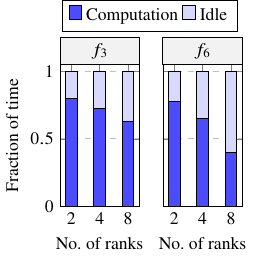}
  }
  \caption{
  Performance of the multi-GPU solver under the round-robin policy as a function of the number of ranks.
(a) Strong scaling for $f_2$ and $f_6$ in $d=6$.\\ 
(b) Computation and idle time fractions for $f_3$ and $f_6$ in $d=6$ at $\tau_{\text{rel}} = 10^{-8}$.}
  \label{fig:scaling-idle}
\end{figure}
Beyond feasibility, we analysed scalability. 
Figure~\ref{fig:scaling} reports the strong scaling of the round-robin redistribution policy for representative integrands at different tolerances. 
Runtimes improved when moving from two to four devices, but beyond this point the benefits flattened or even reversed, with execution times increasing at 8 and 12 GPUs for several functions.
This outcome is not surprising, as adaptive algorithms produce irregular workloads that hinder load balancing at scale.
Three factors explain the observed behaviour: (i) \textit{synchronisation overhead}, since each iteration requires all ranks to align at the global synch point, forcing under-loaded devices to wait for overloaded ones; (ii) \textit{propagation of imbalance}, since the number of subregions exchanged per round is capped, heavily loaded ranks may remain overloaded across multiple iterations; (iii) \textit{bursty and only partially overlapped communication}. Redistribution is asynchronous and overlaps partly with the `evaluate subregions' phase, which dominates the iteration runtime.
However, it is often still too short to cover all transfer times. 
Thus, many messages remain in flight when the global convergence check is reached, and pending communication accumulates at the barrier as idle time.
Idle time increases rapidly with the number of GPUs, in some cases exceeding the useful compute time already at 8 devices (Fig. \ref{fig:idle-time}). 
Although suboptimal in terms of strong scaling, our solver still achieves substantial speedups over PAGANI across the tested functions. 
As shown in Figure~\ref{fig:speedup}, runtimes are reduced by up to an order of magnitude at the same tolerance and dimension. This confirms that, despite limited scalability, the multi-GPU solver provides clear performance gains relative to the state-of-the-art.

\section{Conclusions}
We presented a general-purpose adaptive quadrature strategy for multi-GPU systems, where a decentralised redistribution policy enables highly accurate and fast numerical integration in high dimensions. Although strong scalability is limited by synchronisation and load imbalance, the results demonstrate that multi-GPU adaptivity is essential to broaden the applicability of deterministic quadrature.
Preliminary experiments with more sophisticated load balancing policies have shown potential for reducing idle time, and future work should further investigate this direction to improve the strong scaling behaviour of the current implementation. Another promising direction is the design of more informed initial domain decomposition among GPUs to alleviate the imbalance from the outset.

\bibliographystyle{spmpsci}

\end{document}